\documentclass[twocolumn,preprintnumbers,amsmath,amssymb,prl,superscriptaddress,showpacs]{revtex4}
\usepackage{graphicx}
\usepackage{dcolumn}
\usepackage{bm}
\usepackage{psfrag}
\begin{document}
\renewcommand{\bbox}[1]{{\bm #1}}
\title{Clustering by mixing flows}
\author{Kevin Duncan}
\affiliation{Faculty of Mathematics and Computing,
The Open University, Walton Hall, Milton Keynes, MK7 6AA, England}
\author{Bernhard Mehlig}
\author{Stellan \"Ostlund}
\affiliation{Department of Physics, G\"oteborg University, 41296 Gothenburg, Sweden}
\author{Michael Wilkinson}
\affiliation{Faculty of Mathematics and Computing,
The Open University, Walton Hall, Milton Keynes, MK7 6AA, England}

\begin{abstract}
We calculate the Lyapunov exponents for particles suspended
in a random three-dimensional flow, concentrating on the limit 
where the viscous damping rate is small compared to the inverse correlation
time. In this limit Lyapunov exponents are obtained
as a power series in $\epsilon$,  a dimensionless measure
of the particle inertia. 
Although the perturbation generates an asymptotic series,
we obtain accurate results from a Pad\' e-Borel 
summation. Our results prove that particles suspended in an 
incompressible random mixing flow can show pronounced clustering 
when the Stokes number is large and we characterise two distinct
clustering effects which occur in that limit.
\end{abstract}
% 05.40.-a Fluctuation phenomena, random processes, noise, and Brownian  motion
% 05.60.Cd Classical transport
% 46.65.+g Random phenomena and media
\maketitle
This letter describes the dynamics of particles suspended in a 
randomly moving incompressible fluid which we assume to be mixing:
any given particle uniformly samples configuration space.
At first sight, it seems as if the particles suspended in an
incompressible mixing flow should
become evenly distributed. This indeed happens if the particles are simply
advected by the fluid.
However, it has been noted \cite{Max87} that when the finite 
inertia of the suspended particles is significant, the 
particles can show a tendency to cluster. 

The current %theoretical 
understanding of this remarkable
phenomenon 
refers  to a dimensionless
parameter termed the Stokes number, ${\rm St} = 1/(\gamma\tau)$,
where $\gamma $ is the rate at which the particle velocity 
is damped relative to that of the fluid due to viscous drag, 
and $\tau$ is the correlation time of the velocity of the fluid.
There is a consensus 
\cite{Fes94,Hog01,Sig02,Bec03,Gam04}
that clustering
is most pronounced when 
${\rm St}$ 
is of order unity.

In this letter we 
argue that strong clustering can occur 
when St is large. We show that different clustering mechanisms
compete at large values of St and quantify under which
circumstances clustering occurs.
Before describing our results and outlining how they are derived,
we briefly summarise previous theoretical work on
the clustering of inertial particles in turbulent flows.

This effect was first discussed by Maxey \cite{Max87}: 
he approximated the inertial particle dynamics
by advection in a  \lq synthetic' velocity field
which was obtained as a perturbation of the velocity 
field of the fluid, $\mbox{\boldmath$u$}(\mbox{\boldmath$r$},t)$.
Maxey showed that this synthetic velocity
field has negative divergence when the vorticity of 
$\mbox{\boldmath$u$}(\mbox{\boldmath$r$},t)$
is high or its strain-rate low, and predicted that
particles would have low concentrations in regions
of high vorticity due to this \lq centrifuge effect'.
This effect has been demonstrated in direct numerical
simulation of particles suspended in a fully-developed
turbulent flow \cite{Wan93,Hog01}. 
The theoretical work of Maxey and experimental work
on turbulent flows \cite{Fes94} has emphasised instantaneous
correlations between vortices and particle-density
fluctuations. 

Later work has adapted results on the density statistics
and Lyapunov exponents of purely advective flows 
obtained in \cite{LeJ85,Ber00}: Elperin \cite{Elp96} suggested
combining these results with Maxey's synthetic velocity
field to obtain results for inertial particles; a similar
approach was used in \cite{Pin99,Bal01,Fal02}. These
results are not applicable at large St,
because the perturbation of the velocity field
need not be small when inertial effects are important.

An alternative viewpoint arises from work of Sommerer and Ott
\cite{Som93}, who describe patterns formed by particles floating on a randomly
moving fluid. They characterise the patterns in terms of their
fractal dimension and suggest that the fractal dimension
can be obtained from ratios of Lyapunov exponents
of the particle trajectories using a formula proposed by Kaplan and Yorke \cite{Kap79}.

The argument in \cite{Som93}
extends 
to particles suspended
in turbulent three-dimensional incompressible flows. Consider
the Lyapunov exponents $\lambda_1>\lambda_2>\lambda_3$.
They are rate constants
defined in terms of the time dependence of, respectively,
the length $\delta r$ of a small separation between two trajectories,
the area $\delta {\cal A}$ of a parallelogram spanned by two separation
vectors and the volume $\delta {\cal V}$ of a parallelepiped spanned
by a triad of separations:
\begin{eqnarray}
\label{eq: 2.4}
\lambda_1&=&\lim_{t\to \infty} t^{ -1}\log_{\rm e}(\delta r)
\nonumber \\
\lambda_1+\lambda_2&=&\lim_{t \to \infty}  t^{ -1}\log_{\rm
e}(\delta{\cal A})
\nonumber \\
\lambda_1+\lambda_2+\lambda_3&=&\lim_{t \to \infty} t^{ -1}
\log_{\rm e}(\delta{\cal V})
\ .
\end{eqnarray}
The Kaplan-Yorke estimate for the fractal dimension in a
three-dimensional incompressible flow is determined by the
dimensionless quantity (\lq dimension deficit')
\begin{equation}
\label{eq: 0.0}
\Delta=-(\lambda_1+\lambda_2+\lambda_3)/|\lambda_3|
\ .
\end{equation}
When $\Delta >0$, the Kaplan-Yorke estimate of the dimension is
$d_{\rm H}=3-\Delta$,
and $d_{\rm H}=3$ if $\Delta \le 0$.
Clustering effects are significant if the fractal dimension is
significantly lower than the dimension of space.
This proposition
provides a strong motivation to study
the Lyapunov exponents of the problem.

A third mechanism for clustering 
is the following: 
nothing prevents the infinitesimal volume element $\delta\cal V$ 
from collapsing to zero for an instant of time.
These events correspond
to \lq caustics', where faster moving particles overtake
slower ones. Caustics are associated with the density
of particles on a surface becoming very high, facilitating 
the aggregation of suspended particles. This mechanism
was recently proposed as a cause of clustering
of inertial particles \cite{Wil05}, and is also
mentioned briefly in \cite{Fal02}. 
The significance of this effect is determined by the
rate $J$ at which the infinitesimal
volume element goes through zero for a given triplet of nearby
trajectories.

Which of these three mechanisms is most important?
Maxey's centrifuge effect is weak 
at small $\rm St$, where the particles are simply advected. 
There is a consensus that the effect is also weak for
large St, because the vortices do not
persist for a sufficiently long time to be effective,
implying that significant clustering is only observed
when ${\rm St}\sim 1$.
However, there is at present 
no understanding of what happens at large values of St.
In the following we describe 
quantitative results for the Lyapunov exponents $\lambda_j$, for the dimension
deficit $\Delta$ and for the rate of caustic formation $J$: these
are summarised in 
Fig. 1 {\bf a} - {\bf c}. 

\begin{figure}[b]
\includegraphics[width=7.5cm,clip]{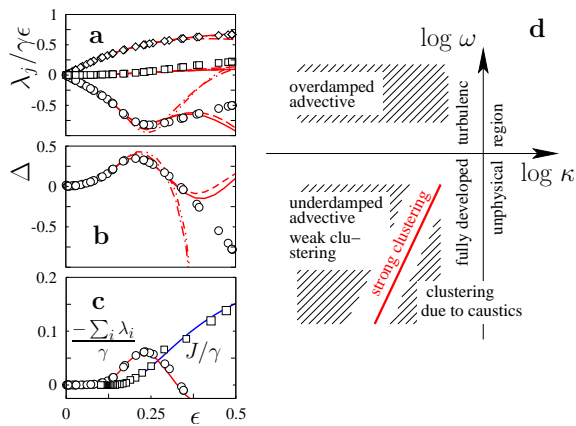}
\caption{\label{fig: 1}
{\bf a} Lyapunov exponents as a function of $\epsilon \sim
\kappa \omega^{-1/2}$: results of simulations of an Euler discretisation
of the linearised
equations of motion (\ref{eq: 2.1}), replacing the elements of ${\bf F}(t)$ 
by random numbers with appropriate statistics, symbols. Also
shown are results based on Pad\'e-Borel summation of (\ref{eq: 6.6}),
with Pad\'e approximants of the orders
$P_{22}$ ($-\cdot\cdot\,-$), $P_{23}$ ($-\cdot-$), $P_{33}$ ($- - -$),
and $P_{34}$ (solid lines).
{\bf b} shows $\Delta$
vs. $\epsilon$ (obtained from (\ref{eq: 0.0}) using the
data in {\bf a}).  {\bf c}
shows $-\sum_i\lambda_i/\gamma$ (circles, solid red line, see {\bf a}
and {\bf b})
compared to the rate of caustic formation determined from simulations 
as described above (squares) and a fit to
$C\,\exp(-S/\epsilon^2)$
with $C = 0.21 $ and $S = 1/8$.
{\bf d} schematic phase diagram in the $\kappa $ -- $\omega $ plane:
note the logarithmic scales.
The red line indicates where
the quantity $\Delta$ is zero. 
Above the red line $\Delta$ is
positive implying clustering.
In the
under- and overdamped advective limits $\Delta$ positive and small (weak
clustering), while below the red line $\Delta$ is negative. In this
regime clustering occurs by caustics.  
}
\end{figure}

Our results show that in order to understand
the clustering effect it is necessary to consider not only
the Stokes number, but an additional dimensionless parameter, $\kappa$,
defined below.
We infer that strong clustering can occur at large
Stokes numbers. Two distinct mechanisms compete (clustering
onto fractal sets versus clustering onto caustics in an otherwise
homogeneous background) and dominate in different
regions of the parameter space.

We model the particles suspended in the fluid flow by
the equation of motion
\begin{equation}
\label{eq: 1.1}
\ddot {\mbox{\boldmath$r$}}= \gamma\big(
\bbox{u}(\mbox{\boldmath$r$},t) -\dot {\mbox{\boldmath$r$}}\big)
\end{equation}
where $\bbox{r} = (r_1,r_2,r_3)$ denotes the position of a particle.
Eq. (\ref{eq: 1.1}) is  appropriate for non-interacting 
spherical particles when the 
Reynolds number of the flow referred to the particle diameter
is small. It is assumed that the radius of the particle and
the molecular mean free path of the fluid are
sufficiently small.  Stokes's formula gives the damping 
rate $\gamma={6\pi a\rho_{\rm f}\nu/{m}}$
where $\nu$, $\rho_{\rm f}$ are respectively the 
kinematic viscosity and density of the fluid, and 
$a$, $m$ are the radius 
and mass of the particle. 
Effects due to the inertia of the
displaced fluid are neglected. This is justified
when the density of the suspended particles is 
large compared to that of the fluid.
We also assume that Brownian diffusion of the particles is 
negligible. 

We now discuss the dimensionless parameters of the problem:
the velocity field is assumed to be characterised
by
its typical velocity $u = \sqrt{\langle
\bbox{u}^2\rangle}$,
by a correlation length $\xi$ and a correlation time $\tau$.
In addition, the interaction of the fluid with the particles
is determined by the damping rate $\gamma$. From these four
quantities we can form two independent dimensionless groups:
a dimensionless velocity, $\kappa = u\tau/\xi$, 
and the dimensionless damping $\omega=\gamma\tau$
(so that $\mbox{St} = \omega^{-1}$).  
The parameter $\kappa$ has been termed \lq Kubo number' \cite{Bri74}. 
It has not been considered before in this context. We argue
that it cannot be large
if $\bbox{u}(\bbox{r},t)$ is to be
a satisfactory model for a solution of the Navier-Stokes equations:
$\tau\le \xi/u$ since disturbances in
the fluid velocity field 
$\bbox{u}(\bbox{r},t)$ are transported by $\bbox{u}(\bbox{r},t)$ itself.

Consider now the particular case of fully-developed turbulence.
In this case, the velocity field exhibits a
power-law energy spectrum, with upper and lower cutoffs
\cite{Fri97}. The smaller length scale is the Kolmogorov length,
which is the size of the smallest vortices generated
by the turbulence. It is given by $(\nu^3/\varepsilon)^{1/4}$,
where $\varepsilon$ is the rate of dissipation per unit mass of fluid.
The Kolmogorov length corresponds to the correlation length $\xi$
in our theory. The corresponding typical velocity $u$ and correlation
time $\tau$ are also determined solely by the same two parameters,
$\varepsilon $ and $\nu$, implying that $\kappa \sim 1$ for fully
developed turbulence. In other situations $\kappa$ can be small.

We now turn to a summary of our results
and outline how they were derived
(details will be published elsewhere).
Linearising the equations of motion (\ref{eq: 1.1}) gives
\begin{eqnarray}
\label{eq: 2.1}
\delta \dot {\mbox{\boldmath$p$}}&=&
-\gamma \delta \mbox{\boldmath$p$}+{\bf F}(t)
\delta \mbox{\boldmath$r$}\,,\quad
\delta \dot {\mbox{\boldmath$r$}}=\delta \mbox{\boldmath$p$}/m
\end{eqnarray}
where $\bbox{p} = m\dot{\bbox{r}}$ is the particle momentum
and ${\bf F}(t)$ is matrix of force gradients:
\begin{equation}
\label{eq: 2.2}
F_{\mu\nu}(t)=\gamma m\, \frac{\partial u_\mu}{\partial r_\nu}
(\mbox{\boldmath$r$}(t),t)
\ .
\end{equation}
We take three
trajectories displaced relative to a reference trajectory by
small increments 
$(\delta \mbox{\boldmath$r$}_\mu,\delta \mbox{\boldmath$p$}_\mu)$,
with $\mu=1,2,3$. 
We introduce a triplet of  orthogonal unit
vectors  ${\bf n}_\nu(t)$ such that  ${\bf n}_1(t)$ is
oriented along $\delta \mbox{\boldmath$r$}_1(t)$, and ${\bf n}_2(t)$ lies
in the plane spanned by
$\big(\delta \mbox{\boldmath$r$}_1(t),\delta \mbox{\boldmath$r$}_2(t)\big)$. 
This determines ${\bf n}_3(t)$ up to a sign which is fixed
by requiring continuity.
We write
${\bf n}_\nu(t) = {\bf O}(t) {\bf n}_\nu(0)$ and 
$\delta \mbox{\boldmath$p$}_\mu(t)= {\bf R}(t)
\,\delta \bbox{r}_\mu(t)$ where ${\bf O}$ is an orthogonal
and ${\bf R}$ a general $3\times 3$ matrix.
We define
the elements of ${\bf F}$ and ${\bf R}$ transformed
to the moving basis:
\begin{equation}
\label{eq: 2.9}
F'_{\mu\nu}(t)\!=\!{\bf n}_\mu (t)\!\cdot\!{\bf F}(t){\bf
n}_\nu(t)\,,\,\,
R'_{\mu\nu}(t)\!=\!{\bf n}_\mu (t)\!\cdot\!{\bf R}(t){\bf n}_\nu(t)
\end{equation}
and find the following equation of motion for ${\bf R}'$
\begin{equation}
\label{eq: 2.25}
\dot {{\bf R}'} = -\gamma {\bf R}' -\frac{1}{m} {{\bf R}'}^2
+[{\bf R}',{\bf O}^+ \dot{\bf O}]
+{\bf F}'\,.
\end{equation}
The elements
of ${\bf O}^+ \dot{\bf O}$ are given by 
\begin{equation}
\label{eq: 2.26}
{\bf O}^+ \dot{\bf O}=\frac{1}{m}
\left( \begin{array}{lcr}
0       &-R_{21}^\prime&-R_{31}^\prime\\
R_{21}' &      0 &-R_{32}'\\
R_{31}' & R_{32}'& 0 
\end{array}\right)\,.
\end{equation}
We find that the Lyapunov exponents are equal to 
the long-time average of
the diagonal elements of ${\bf R}'$
\begin{eqnarray}
\label{eq: 2.20}
\lambda_1&=&\langle R'_{11}\rangle/m\,,\,\,
\lambda_2=\langle R'_{22}\rangle/m\,,\,\,
\lambda_3=\langle R'_{33}\rangle/m\,.
\end{eqnarray} 
Eqs. (\ref{eq: 2.25}) and (\ref{eq: 2.26})  for ${\bf R}'$ can be simplified
when the correlation time of the velocity field is sufficiently
short, $\omega \ll 1$, 
assuming that the amplitude of the random force is sufficiently 
small, $\kappa \ll 1$. In this limit  ${\bf F}'$ 
behaves as a 
white-noise signal, and (\ref{eq: 2.25}) reduces to a 
system of Langevin equations. We label the dynamical variables
by a single index $i = 3(\mu-1)+\nu$ and scale the Langevin
equations for $R'_i$ to
dimensionless form 
\begin{equation}
\label{eq: 3.14}
{\rm d}x_i=-\Big(x_i+\epsilon \sum_{j=1}^9\sum_{k=1}^9 V^i_{jk}
x_j x_k\Big){\rm d}t'+{\rm d}w_i
\end{equation}
Here $t'=\gamma t$, $x_i=\sqrt{{\gamma/{D_1}}}R_i'$, and
$\langle {\rm d}w_i{\rm d}w_j\rangle=2D_{ij}{\rm d}t'$. 
The elements $D_{ij}$ of the diffusion matrix $\bf D$ are
given by
\begin{equation}
\label{eq: 3.7}
D_{ij}={\textstyle{\frac{1}{2}}}\int_{-\infty}^\infty{\rm d}t\,
\langle F_i'(t) F_j'(0)\rangle \ .
\end{equation}
The coefficients $V^i_{jk}$ are determined by the 2nd and 3rd
terms on the rhs of (\ref{eq: 2.25}). The dimensionless 
parameter
\begin{equation}
\label{eq: 3.13}
\epsilon={D_{11}^{1/2}/({m\gamma^{3/2}}})\sim \kappa \omega^{-1/2}
\end{equation}
is a measure of the inertia of the particles: it is proportional to $a$
and therefore to $m^{1/3}$. Thus we obtain all three
Lyapunov exponents from the expectation values of variables in a system
of Langevin equations. Earlier work has  obtained the largest
Lyapunov exponents for various problems using Langevin equations
\cite{Hal65,Pit02,Meh04}.

The elements of ${\bf D}$ are determined by 
the fluctuations of the velocity field.
We assume that the latter is incompressible,
but for reasons explained below  we add a small compressible component:
$\mbox{\boldmath$u$}= \mbox{\boldmath$\nabla$} \wedge \mbox{\boldmath$A$}
+\bbox{\nabla}\delta A_0$.
The fields $A_\mu(\mbox{\boldmath$r$},t),\,\mu=1,2,3$ are taken
to be homogeneous in space and time, and 
isotropic in space. Their correlations are determined by
$\langle A_\mu(\mbox{\boldmath$r$}+\mbox{\boldmath$R$},t_0+t)
A_\nu(\mbox{\boldmath$r$}_0,t_0)\rangle
=\delta_{\mu\nu} \,C(|\bbox{r}-\bbox{r}'|/\xi,|t-t'|/\tau)$.
The field $\delta A_0$ is statistically independent
of $A_\mu$, has the same correlation function, and 
in the end the limit $\delta A_0\rightarrow 0$ is taken.

The Langevin equations (\ref{eq: 3.14}) are equivalent
to a Fokker-Planck equation whose stationary solution $P(\bbox{x})$ 
determines the Lyapunov exponents.
In the limit of $\epsilon\rightarrow 0$
the latter  is Gaussian
\begin{equation}
\label{eq: 10.3}
P_0(\bbox{x})\propto \exp(-{\textstyle{\frac{1}{2}}}\bbox{x}\cdot 
{\bf D}^{-1}\,\bbox{x})
\equiv \exp[-\Phi_0(\bbox{x})]\,.
\end{equation}
This suggests transforming the Fokker-Planck
operator so that its $\epsilon=0$ limit 
is transformed into a harmonic oscillator.  This is achieved
by introducing $Q(\bbox{x})=\exp[\Phi_0(\bbox{x})/2]P(\bbox{x})$.  
The steady-state Fokker-Planck equation can be written as
$(\hat{H}_0+\epsilon \hat H_1)|Q) = 0$, where we have represented
the function $Q(\bbox{x})$ by a \lq ket vector' $\vert Q)$.
The operator $\hat H_0$  is the Hamiltonian for nine uncoupled
harmonic oscillators
\begin{equation}
\label{eq: 10.16}
\hat{H}_0=-\sum_{i=1}^{9}  \hat a^+_i \hat a_i
\end{equation}
where the $\hat a^+_i$ and $\hat a_i$ are, respectively, the
creation and annihilation operators for the degree of freedom 
labelled by $i$ (satisfying
$[\hat a_i,\hat a_j^+]=\delta_{ij}\hat I$). The non-Hermitean
perturbation $\hat H_1$ can be expressed in terms of 
the eigenvalues $\omega_i$ of ${\bf D}$ and the elements $J_{ij}$ 
of an orthogonal matrix ${\bf J}$ satisfying 
${\bf D}={\bf J}\mbox{\boldmath$\Omega$}{\bf J}^{-1}$, with 
$\mbox{\boldmath$\Omega$}={\rm diag}(\omega_i)$:
\begin{eqnarray}
\label{eq: 10.24}
\hat{H}_1&=&\sum_{ijk} H^{(1)}_{ijk}\,
\hat a^+_i(\hat a^+_j+\hat a_j)(\hat a^+_k+\hat a_k)
\nonumber \\
H^{(1)}_{ijk}&=&
\sqrt{\omega_j\omega_k/\omega_i}\,\, \sum_{lmn} V^l_{mn}J_{il}J_{mj}J_{nk}
\ . 
\end{eqnarray}
Regularisation is needed since
one eigenvalue vanishes in the limit of $\delta A_0\rightarrow 0$.
We determine $|Q)$ by perturbation theory in $\epsilon$.
Given $|Q)$, the  Lyapunov exponents are obtained as
$\lambda_1 = \gamma\epsilon \langle x_1\rangle$,
$\lambda_2 = \gamma\epsilon \langle x_5\rangle$, 
$\lambda_3 = \gamma\epsilon \langle x_9\rangle$, and
\begin{eqnarray}
\label{eq: 6.1}
\langle x_i\rangle&=&
\frac{1}{(\Phi_{\bf 0}\vert Q)}
\sum_j J_{ij}\sqrt{\omega_j}(\Phi_{\bf 0}\vert \hat a_j+\hat a^+_j\vert Q)
\end{eqnarray}
where $\vert \Phi_0 )$ denotes the null eigenvector of 
$\hat H_0$. 
From (\ref{eq: 6.1}) we obtain series expansions 
in the form
\begin{eqnarray}
\lambda_1/\gamma &=& 3\epsilon^2-29\epsilon^4 +564\epsilon^6\nonumber\\
&&-14977\epsilon^8+488784\epsilon^{10}-18670570\epsilon^{12}+\cdots
\nonumber\\
\lambda_2/\gamma &=&
8\epsilon^4-459/2\,\epsilon^6+14281/2\,\epsilon^8
\label{eq: 6.6}\\
&&-757273/3\,\epsilon^{10}
+361653709/36\,\epsilon^{12}+\cdots
\nonumber \\
\lambda_3/\gamma &=& -3\epsilon^2-9 \epsilon^4-789/2\,\epsilon^6
-5787/2\,\epsilon^8
\nonumber\\
&&-895169/3\,\epsilon^{10}-101637719/36\,\epsilon^{12}+\cdots\,.
\nonumber
\end{eqnarray}
Note that only even powers of $\epsilon$ contribute, and that
all coefficients are rational numbers.
Eq. (\ref{eq: 6.6}) is the main result of this letter. The
expansion is valid in the underdamped limit $\omega \ll 1$
when $\kappa \ll 1$.

The coefficients in (\ref{eq: 6.6}) exhibit rapid growth 
typical of an asymptotic series \cite{Boy99}.
We have attempted to sum the series (\ref{eq: 6.6})
using Pad\'e-Borel summation \cite{Boy99}
\begin{equation}
\label{eq:boreltransform}
\lambda_j/\gamma \sim \mbox{Re}\int_C {\rm d}t\,  {\rm e}^{-t} 
\, \sum_{l=1}^{\rm l_{\rm max}} 
\frac{c_l^{(j)}}{l!} \epsilon^{2l}
\end{equation}
where $c_l^{(j)}$ are the coefficients of (\ref{eq: 6.6}) and
$l_{\rm max}=7$ is the number of nonzero coefficients available
for each $\lambda_j$.
The sum in the integrand 
is approximated
by Pad\'{e} approximants \cite{Ben78} of order $n$, namely $P_{n}^n$ or $P_{n+1}^n$
with $n \leq [l_{\rm max}/2]$.
The integration path in (\ref{eq:boreltransform}) is taken to be a 
ray in the upper right
quadrant in the complex plane.

Results of Pad\'e-Borel summations of the series for $\lambda_j$ 
are shown in Fig. \ref{fig: 1}{\bf a} and converge to results
of numerical simulations provided $\epsilon$ is not too large.
For $\lambda_2$ numerical evidence indicates the presence of additional
non-analytical contributions
not captured by the Pad\'e-Borel summation.
The results of Fig. 1{\bf a} allow us to determine the quantity
$\Delta$ defined in eq. (\ref{eq: 0.0}). The result is shown in Fig. 1{\bf b}. 
We find that $\Delta$ is maximal
for $\epsilon \approx 0.21$ and positive (indicating
clustering onto a fractal set) for $0 < \epsilon < 0.33$.
The red line in Fig. \ref{fig: 1}{\bf d}, $\epsilon \sim  \kappa \omega^{-1/2} =
\mbox{const}.$, indicates schematically
where $\Delta$ is zero. Above the red line
$\Delta$ is always positive, but tends to zero for small $\epsilon$
as $\Delta = 10\epsilon^2 \sim \kappa^2/\omega$. 
In the limit of $\epsilon\rightarrow 0$ the 
dynamics becomes advective (despite being underdamped):
to lowest order in $\epsilon$ our results
coincide with those for purely advective
flow \cite{LeJ85}. 

We now turn to the rate $J$ of caustic formation.
It is the rate at which
$\delta V(t) = (\delta\bbox{r}_1(t)\wedge \delta\bbox{r}_2(t))\cdot
\delta\bbox{r}_3(t)$ goes through zero. Since $\delta\bbox{p}_\mu$
typically remain bounded, caustics correspond to instances
where the elements of the third column of ${\bf R}'$ go to $-\infty$
and reappear at $\infty$.
The rate at which these events occur is given by the 
escape rate of the Langevin process (\ref{eq: 3.14}) to infinity. 
It is expected
\cite{Wil05} to have a non-analytic dependence on $\epsilon$, of the
form $C\,\exp(-S/\epsilon^2)$, as demonstrated in  Fig. 1{\bf c}.
In this panel, $J/\gamma$ is compared to
$-(\lambda_1+\lambda_2+\lambda_3)/\gamma$. We see that 
caustics are very rare when $\epsilon \ll 1$,
but frequent when $\epsilon$ is large and they
are the only clustering mechanism when $\epsilon >0.33$.

Finally, we comment on the relation between our results
and earlier works (cited above), which suggest that clustering only
occurs for $\omega \approx 1$ (with the value of $\kappa$
unspecified). It must
be emphasised that the earlier quantitative theoretical results on clustering
are confined to the overdamped limit $\omega \gg 1$, where inertial
effects are small: for purely advective flow there is no clustering
($\Delta=0$ and $J=0$). Inertial effects were incorporated by
Elperin and others \cite{Elp96,Pin99,Bal01}, using
Maxey's perturbative correction to the velocity. 
Their results are valid only for the limit $\omega \gg 1$,
and are distinct from our series expansions  (\ref{eq: 6.6}): 
this is most easily seen by calculating corrections to $\Delta$
in this overdamped limit. 
We find that 
$\Delta \sim\kappa^2/\omega^2$ implying that
clustering effects are small in this regime. In the underdamped
regime, by contrast, we obtained $\Delta \sim\kappa^2/\omega$
which can be of order unity.

The results of this letter are summarised schematically in figure 1{\bf d}. 
First, at small $\kappa$, strong clustering occurs 
in the region indicated, above the line $\epsilon \sim \kappa
\omega^{-1/2} = 0.33$. Second, since the dimension deficit
$\Delta$ is positive in this regime, the reasoning of Sommerer and Ott
\cite{Som93}
indicates that the particles cluster on a fractal.
Third, as  $\epsilon\rightarrow 0$ the dynamics becomes
advective.
In this limit the dimension deficit $\Delta$ and the
rate of caustic formation $J$ vanish: 
particles advected in an incompressible flow remain
uniformly distributed.
Fourth, when $\epsilon > 0.33$  we find that
the dimension deficit $\Delta$ is negative implying that
do not lie on a fractal. They are however not homogenously
distributed: in this regime 
particles cluster because they are brought into close contact
by caustics.

\end{document}